\documentclass[fleqn,11pt]{article}
\usepackage{url}
\usepackage{algorithmic}
\usepackage{amsfonts}
\usepackage{enumerate}

\usepackage{color}
\usepackage{mathptmx}
\usepackage{graphics}
\usepackage{epsf}

\newcommand{\bfe}[1]{\begin{bfseries}\emph{#1}\end{bfseries}\index{#1}}
\newcommand{\ES}{\mbox{$\emptyset$}}

\newcommand{\myra}{\mbox{$\:\rightarrow\:$}}

\newcommand{\La}{\mbox{$\:\Leftarrow\:$}}
\newcommand{\Ra}{\mbox{$\:\Rightarrow\:$}}
\newcommand{\tra}{\mbox{$\:\rightarrow^*\:$}}

\newcommand{\A}{\mbox{$\ \wedge\ $}}

\newcommand{\sse}{\mbox{$\:\subseteq\:$}}

\newcommand{\LL}{\mbox{$\ldots$}}

\newcommand{\C}[1]{\mbox{$\{{#1}\}$}}           

\newcommand{\NI}{\noindent}
\newcommand{\HB}{\hfill{$\Box$}}
\newcommand{\VV}{\vspace{5 mm}}

\newcommand{\II}{\vspace{2 mm}}

\newcommand{\szkew}[1]{\relax \setbox0=\hbox{\kern -24pt $\displaystyle#1$\kern 0pt }%
\box0}
{\catcode`\@=11 \global\let\ifjusthvtest@=\iffalse}

\newcounter{oldmycaption}

\newcommand{\Proof}{\NI
                    {\bf Proof.}\ }

\newtheorem{theorem}{Theorem}[section]
\newtheorem{defined}[theorem]{Definition}
\newenvironment{definition}{\begin{defined} \rm}{\end{defined}}
\newtheorem{exa}[theorem]{Example}
\newenvironment{example}{\begin{exa} \rm}{\end{exa}}

\newtheorem{lemma}[theorem]{Lemma}
\newtheorem{corollary}[theorem]{Corollary}

\newtheorem{note}[theorem]{Note}

\newtheorem{exe}{Exercise}

\newtheorem{pro}{Problem}

\newcounter{symbol}
\newcommand{\indexsyma}[1]%
{\stepcounter{symbol}\index{zzz1 \thesymbol @\protect#1}}
\newcommand{\indexsymb}[1]%
{\stepcounter{symbol}\index{zzz2 \thesymbol @\protect#1}}
\newcommand{\indexsymc}[1]%
{\stepcounter{symbol}\index{zzz3 \thesymbol @\protect#1}}
\newcommand{\indexsymd}[1]%
{\stepcounter{symbol}\index{zzz4 \thesymbol @\protect#1}}
\newcommand{\indexsyme}[1]%
{\stepcounter{symbol}\index{zzz5 \thesymbol @\protect#1}}

\newcommand{\snet}{\mathcal{S}}

\title{Social Networks with Competing Products\footnote{A preliminary version of this paper appeared as \cite{AM11}.}}

\author{Krzysztof R. Apt \\
\emph{CWI, Amsterdam, the Netherlands and University of Amsterdam} \\
\and
Evangelos Markakis \\
\emph{Athens University of Economics and Business, Athens, Greece}
}
\date{}
\begin{document}
\maketitle


\begin{abstract}
We introduce a new threshold model of social networks, in which the
nodes influenced by their neighbours can adopt one out of several
alternatives.  We characterize social networks
for which adoption of a product by the
whole network is possible (respectively necessary) and the ones
for which a unique outcome is guaranteed.
These characterizations directly yield
polynomial time algorithms that allow us to determine whether a
given social network satisfies one of the above properties.

We also study algorithmic questions for networks without unique
outcomes. We show that the problem of determining 
whether a final network exists in which all nodes adopted some product
is NP-complete. In turn, we also resolve the complexity of the problems of determining whether
a given node adopts some (respectively, a given) product in 
some (respectively, all) network(s).

Further, we show that the problem of computing the minimum possible
spread of a product is NP-hard to approximate with an approximation
ratio better than $\Omega(n)$, in contrast to the maximum spread,
which is efficiently computable.  Finally, we clarify that some of the
above problems can be solved in polynomial time when there are only
two products.
\end{abstract}

\section{Introduction}

\subsection{Background}

Social networks have become a huge interdisciplinary research area with
important links to sociology, economics, epidemiology, computer
science, and mathematics.  A flurry of numerous articles, notably the
influential \cite{Mor00}, and recent books, see
\cite{Cha04,Goy07,Veg07,Jac08,EK10} shows the growing relevance of
this field. It deals with such diverse topics as epidemics, spread of
certain patterns of social behaviour, effects of advertising, and
emergence of `bubbles' in financial markets.

A large part of research on social networks focusses on the problem of
\emph{diffusion}, that is the spread of a certain event or information
over the network, for example becoming infected or adopting a
given product. In the remainder of the paper, we will use as a running example the
adoption of a product, which is being marketed over a social network.

Two prevalent models have been considered for capturing diffusion: the
{\em threshold models} introduced in~\cite{Gra78} and~\cite{Sch78} and
the {\em independent cascade models} studied in~\cite{GLM01}. In the
threshold models, each node $i$ has a threshold $\theta(i) \in (0,1]$
and it decides to adopt a product when the total weight of incoming
edges from nodes that have already adopted a product exceeds the
threshold. In a special case a node decides to adopt a new product if
at least the fraction $\theta(i)$ of its neighbours has done so. In some cases 
the threshold may also depend on the specific product under consideration.
In the cascade models, each node that adopts a product can activate
each of his neighbours with a certain probability and each node has
only one chance of activating a neighbour\footnote{For the case of a
single product, and when thresholds are assumed to be random variables,
the two models have been proved to be equivalent in the sense that
they produce the same distribution on outcomes~\cite{KKT03}.}.

Most of research has focussed on the situation in which the players
face the choice of adopting a specific product or not. The algorithmic
problem of choosing an initial set of nodes so as to maximize the
adoption of a given product were studied
initially in~\cite{DR01} and~\cite{KKT03}. Certain variants and generalizations 
of this problem were also studied in several publications
that followed, e.g., \cite{Che09,GK11,MR07}.

When studying social networks from the point of view of adopting new
products it is natural to lift the restriction of one product.
One natural example of such a situation is when users can adopt one out of
several competing products (for example providers of mobile
telephones). Then, because of lower subscription costs, each owner of
a mobile telephone naturally prefers that his friends choose the same
provider. Another example is when children have to choose a secondary
school. Here, again, children prefer to choose a school which their
friends will choose, as well. Also, in discussions preceding voting in
a small institution, for instance for the position of a chairman of a
club, preferences announced by some club members may influence the
votes cast by their friends.

What is common in these situations is that the number of choices is
small in comparison with the number of agents and the outcome of the
adoption process does not need to be unique.  Indeed, individuals with
a low 'threshold' can adopt any product a small group of their friends
adopts. As a result this model leads to different considerations than
the ones mentioned above.

Social networks in the presence of multiple products have
been studied in a number of recent papers.
In the presence of multiple products, diffusion has been investigated
recently for the cascade model in~\cite{BKS07,CNWZ07,KOW08}.
In~\cite{KOW08} a special case of the cascade model is studied and
NP-hardness results are obtained on finding the best set of
influential nodes in the presence of another competing product.
In~\cite{BKS07} the authors also study a generalization to a cascade
model with multiple products and provide approximation algorithms for
the problem of maximizing the influence of a product given the initial
adopters of the other products. Finally, in~\cite{CNWZ07}, the authors
provide approximation algorithms for certain variants of the problem
with two products.

For threshold models, an extension to two products has
been recently proposed in~\cite{BFO10}, where the authors examine whether the
algorithmic approach of~\cite{KKT03} can be extended. 
Algorithms and hardness of approximation results
are provided for certain variants of the diffusion process. In line
with~\cite{KKT03}, the authors of~\cite{BFO10} also assume that the
threshold of each node is a random variable and the goal is to
maximize the expected spread.

Game theoretic aspects have also been considered in the case of two or more
products. In particular, the behavior of best response dynamics in
infinite graphs is studied in~\cite{Mor00}, when each node has to
choose between two different products. An extension of this model is
studied in~\cite{IKMW07} with a focus on notions of compatibility and
bilinguality, i.e., having the option to adopt both products at an
extra cost so as to be compatible with all your neighbours. Other game theoretic approaches have also been recently considered, by viewing 
the firms that market their products as strategic agents, see \cite{AFPT10,GK12,TAM12}.

\subsection{Contributions}

We study a new model of a social network in which
nodes (agents) can choose out of \emph{several} alternatives
\emph{and} in which various outcomes of the adoption process are
possible. Our model combines a number of features present in 
various models of networks and is a natural next step in this line 
of research.

It is a threshold model and we assume that the threshold of a node is
a fixed number as in~\cite{Che09} (and unlike \cite{KKT03,BFO10},
where they are random variables).  This is in contrast to Hebb's model
of learning in networks of neurons, the focus of which is on learning,
leading to strengthening of the connections (here thresholds). In our
context, the threshold should be viewed as a fixed `resistance level'
of a node to adopt a product. Contrary to the SIR model, see, e.g.,
\cite{Jac08}, in which a node can be in only two states, in our model
each node can choose out of several states (products). We also allow
that not all nodes have exactly the same set of products to choose
from, e.g. due to geographic or income restrictions some products may
be available only to a subset of the nodes. If a node changes its
state from the initial one, the new state (that corresponds to the
adopted product) is final, as is the case with most of the related
literature.

Our work consists of two parts. In the first part
(Sections~\ref{sec:reachable}, \ref{sec:unavoidable},
\ref{sec:unique}) we study three basic problems concerning this model,
motivated by the nondeterministic character of the adoption process.
In particular, we find necessary and sufficient conditions for
determining whether

\begin{itemize}
\item a specific product will possibly be adopted by all nodes,

\item a specific product will necessarily be adopted by all nodes,

\item the adoption process of the products will yield a unique outcome.
\end{itemize}

For each of these questions, we obtain a characterization with respect
to properties of the underlying graph.  Furthermore, our
characterizations yield efficient algorithms for solving each problem.
We also identify a natural class of social networks that yield a
unique outcome.

In the second part (Section~\ref{sec:adoption-analysis}) we
investigate the complexity of various other algorithmic problems
concerning the adoption process. We start with the problem of
determining whether, given an initial network, a final network exists
in which all nodes adopted a product.  Then we move on to questions
regarding the behaviour of a given node in terms of adopting a given
product or some product in some (respectively, all) network(s).  
We also study the problems of computing the minimum (respectively, maximum)
possible spread of a product.

We resolve the complexity of all these problems. Some of them
turn out to be efficiently solvable, whereas the remaining
ones are either co-NP-complete or have strong inapproximability
properties.  We also show that some, but not all, of these problems
can be solved in polynomial time when there are only two products.

Finally, in Section~\ref{sec:structural} we explain how one can
transform social networks into ones that are in some sense simpler, at
the cost of addition of new nodes. 
These transformations clarify the conciseness hidden in the initial
definition and relate it to the one used in \cite{AM11}, in which the
threshold functions were product independent.

\section{Preliminaries}

Assume a fixed weighted directed graph $G = (V, E, w)$ (with no parallel
edges and no self-loops), with $n = |V|$ and $w_{ij}\in [0, 1]$ being
the weight of edge $(i, j)$. 
In our proposed algorithms we shall assume
that we are given the adjacency matrix representation of the graph.
Some of our algorithms use the adjacency lists representation, 
which can be easily obtained from the adjacency matrix
in time $O(n^2)$.

Given a node $i$ of $G$ we denote by
$N(i)$ the set of nodes from which there is an incoming edge to $i$.
We call each $j \in N(i)$ a \bfe{neighbour} of $i$ in $G$.
We assume that for each node $i$ such that $N(i) \neq \ES$, $\sum_{j
\in N(i)} w_{ji} \leq 1$.

Let $P$ be a finite set of alternatives, that we call from now on
\bfe{products}.  By a \bfe{social network} (from now on, just
\bfe{network}) we mean a tuple $(G,P,p,\theta)$, where $p$ assigns to
each agent $i$ a non-empty set of products $p(i) \sse P$ from which it
can make a choice. For $i \in V$ and $t \in p(i)$ the \bfe{threshold
  function} $\theta$ yields a value $\theta(i,t) \in (0,1]$. The
threshold $\theta(i,t)$ should be viewed as agent $i$'s resistance
level to adopt product $t$. In some cases, the threshold function may
not depend on the product $t$. We will call such functions
\bfe{product independent} and we will then use $\theta(i)$ instead of
$\theta(i, t)$ to denote the resistance of agent $i$.

The idea is that each node $i$ is offered a non-empty set
$p(i)$ of products from which it can make its choice.  If
$p(i)$ is a singleton, say $p(i) = \{t\}$, the node adopted the
product $t$. Otherwise it can adopt a product $t$ if the total weight of
incoming edges from neighbours that have already adopted $t$ is at
least equal to the threshold $\theta(i,t)$. To formalize the problems that we want to study, we need first to introduce a number of
notions.  Since $G,P$ and $\theta$ are fixed, we often identify each
network with the function $p$.

Consider a binary relation $\myra$ on networks.  Denote
by $\tra$ the reflexive, transitive closure of $\myra$.  We call a
reduction sequence $p \tra p'$
\bfe{maximal} if for no $p''$ we have $p' \myra
p''$. In that case we will say that $p'$ is a \bfe{final} network, given the initial network $p$.


\begin{definition}
Assume an initial network $p$ and  a network
$p'$. We say that

\begin{itemize}

\item  $p'$ is \bfe{reachable} (from $p$) if $p \tra p'$,

\item $p'$ is \bfe{unavoidable} (from $p$)
if for all maximal sequences of reductions
$p \tra p''$ we have $p' = p''$,

\item $p$ admits a \bfe{unique outcome}
if some network is unavoidable from $p$.
\HB
\end{itemize}
\end{definition}

So a network is reachable if it can be reached by some sequence
of $\myra$ reductions that starts with $p$, and it is unavoidable if
it is reachable by a maximal sequence of reductions and a unique
outcome of the initial network $p$ exists.

From now on we specialize the relation $\myra$.  Given a social
network $p$ when $N(i) \neq \ES$ 
we use the abbreviation $A(t,i)$ (to stand for "adoption condition
for product $t$ by node $i$") for the condition
\[
\sum_{j \in N(i)\mid p(j) = \{t\} } w_{ji} \geq \theta(i,t)
\]
and stipulate without loss of generality that $A(t,i)$ holds when $N(i) = \ES$.

\begin{definition}
\mbox{} \\[-4mm]

\begin{itemize}
\item
We write $p_{1} \myra p_{2}$ if $p_2 \neq p_1$ and
for all nodes $i$, if $p_2(i) \neq p_1(i)$, then $|p_1(i)|  \geq 2$ and
for some $t \in p_1(i)$
\[
p_{2}(i) = \{t\} \mbox{ and $A(t,i)$ holds in $p_1$}.
\]

\item We say that node $i$ in a network $p$
\begin{itemize}
\item
\bfe{adopted product $t$} if $p(i) = \{t\}$,

\item \bfe{can adopt product $t$} if
\[
\mbox{$t \in p(i) \A |p(i)| \geq 2 \A A(t,i)$.}
\]
\HB
\end{itemize}
\end{itemize}
\end{definition}

In particular, a node with no neighbours and more than one available product
can adopt any product
that is a possible choice for it. As with most of the literature on diffusion models, we also assume that an adoption decision is final. Once a node decides
to adopt a product, it cannot cancel its decision or switch later to another product. In a follow up paper, \cite{SA12}, a game-theoretic framework is considered in which the agents' decisions are not final in the sense that they can switch
to a more attractive product than the current one.

So $p_{1} \myra p_{2}$ holds if
\begin{itemize}
\item any node that adopted a product in $p_2$ either adopted it in
  $p_1$ or could adopt it in $p_1$,

\item at least one node could adopt a product in $p_1$ and adopted it in $p_2$,

\item the nodes that did not adopt a product in $p_2$ did not change their product sets.
\end{itemize}

Note that each modification of the function $p$ results in assigning to a node $i$
a singleton set. So if $p_{1} \tra p_{2}$, then for all nodes
$i$ either $p_{2}(i) = p_{1}(i)$ or $p_{2}(i)$ is a singleton set.

One can naturally incorporate in the reduction process an elimination of products that cannot be 
adopted. More precisely, given a network $p$ and a node $i$, suppose that for some product $t \in p(i)$
\[
\sum_{j \in N(i) \mid t \in p(j)} w_{ji} <  \theta(i,t).
\]
Then $t$ can never be adopted by node $i$. Call such a product $t$
\bfe{infeasible} for $i$.  Clearly, at each state of the reduction
process all infeasible products can be discarded from the
corresponding product sets $p(i)$ without affecting the adoption
process (and hence without affecting our results).  We therefore allow
only adoption steps in our model and no elimination steps.

One of the questions we are interested in is whether a product $t$ can spread to the whole network.
We will denote this final network by $[t]$, where $[t]$ denotes the constant function $p$ such
that $p(i) = \{t\}$ for all nodes $i$.

Below, given a network $(G,P,p,\theta)$ and a product $t \in P$ we denote
by $G_{p,t}$ the weighted directed graph obtained from $G$ by removing from it
all edges to nodes $i$ with $p(i) = \{t\}$. So in $G_{p,t}$ for all such nodes $i$
we have $N(i) = \ES$ and for all other nodes the set of neighbours in $G_{p,t}$
and $G$ is the same.

If each weight  $w_{ji}$ in the considered  graph
equals $\frac{1}{|N(i)|}$, then we call the corresponding network \bfe{equitable}.
So in equitable networks 
the adoption condition $A(t,i)$ holds if at least the fraction $\theta(i,t)$ of 
the neighbours of $i$ adopted in $p$ product $t$.

\begin{example}
\label{ex:nets}

As an example consider the equitable networks in Figure
\ref{fig:soc}, where $P = \{t_1, t_2\}$ and where we mention next to
each node the set of products available to it. 
We assume in this example that the threshold function does not depend on the product argument, hence we omit it.

\begin{figure}[htbp]
\begin{center} \ \setlength{\epsfxsize}{6cm}
\epsfbox{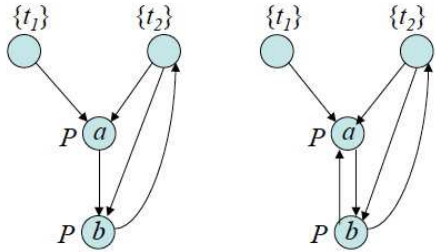}

\end{center}
\caption{Two examples of social networks} \label{fig:soc}
\end{figure}

In the first network, if $\theta(a) \leq \frac{1}{2}$, then the
network in which each node apart from the one on the top left adopts product $t_2$ is reachable, though not unavoidable. 
It is no longer a reachable network if $\theta(a) > \frac{1}{2}$. 
In that case the initial network admits a unique outcome. In this unique outcome, node $b$ adopts product $t_2$ if and only if
$\theta(b) \leq \frac{1}{2}$.


For the second network the following more elaborate case
distinction lists the possible values of $p$ in the final reachable
networks.

\[
\begin{array}{ll}
\theta(a) \leq \frac{1}{3}  \A \theta(b) \leq \frac{1}{2}& : (p(a) = \{t_1\} \vee p(a) = \{t_2\}) \A \\
                                                         & \ \ (p(b) = \{t_1\} \vee p(b) = \{t_2\}) \\[1mm]
\theta(a) \leq \frac{1}{3}  \A \theta(b) > \frac{1}{2} & : (p(a) = \{t_1\} \A p(b) = P) \: \vee \\
                                                       & \ \ (p(a) = p(b) = \{t_2\}) \\[1mm]
\frac{1}{3} < \theta(a) \leq \frac{2}{3}  \A \theta(b) \leq \frac{1}{2} & : p(a) = p(b) = \{t_2\} \\[1mm]
\frac{1}{3} < \theta(a) \A \theta(b) > \frac{1}{2} & : p(a) = p(b) = P \\[1mm]
\frac{2}{3} < \theta(a) \A \theta(b) \leq \frac{1}{2} & : p(a) = P \A p(b) = \{t_2\} 
\end{array}
\]

In particular, when $\frac{1}{3} < \theta(a) \leq \frac{2}{3}$
and $\theta(b) \leq \frac{1}{2}$
there is a unique reduction sequence in which 
first node $b$ adopts product $t_2$ \emph{followed} by node $a$ adopting $t_2$.
\HB
\end{example}

\section{Reachable outcomes}
\label{sec:reachable}

We start with providing necessary and sufficient conditions for a product to be
adopted by all nodes. We shall need the following notion.

\begin{definition}
\label{def:q}
Given a weighted directed graph $G$, a threshold function $\theta$ and a product $t$,
we will say that $G$ is
\bfe{($\theta,t$)-well-structured} if for some function $\mathtt{level}$ that maps
nodes to natural numbers, we have that for all nodes $i$ such that $N(i) \neq \ES$
\begin{equation}
\label{equ:theta}
\sum_{j \in N(i) \mid \mathtt{level}(j) < \mathtt{level}(i)} w_{ji} \geq \theta(i,t).
\end{equation}
\end{definition}

In other words, a weighted directed graph is ($\theta,t$)-well-structured if levels
can be assigned to its nodes in such a way that for each node $i$
such that $N(i) \neq \ES$, the
sum of the weights of the incoming edges from lower levels is at least
$\theta(i,t)$.  We will often refer to the function $\mathtt{level}$ as a
\bfe{certificate} for the graph being ($\theta,t$)-well-structured. Note
that there can be many certificates for a given graph. 
Note also that ($\theta,t$)-well structured graphs can have cycles. For instance, it is
easy to check that for every product $t \in P$
the second network in Figure \ref{fig:soc} is
($\theta,t$)-well structured when $\theta(i) \leq \frac{1}{3}$ for every node $i$.


We provide now a structural characterization of graphs that allow
products to spread to the whole graph, given the threshold function
$\theta$. This will allow us to efficiently determine whether a given
product can spread to the whole network.

\begin{theorem} \label{thm:res1}
Assume a network $(G,P,p,\theta)$ and a product $top \in P$.
The network $(G,P,[top],\theta)$ is reachable from $(G,P,p,\theta)$ iff
\begin{itemize}

\item for all $i$, $top \in p(i)$,

\item $G_{p,top}$ is ($\theta,top$)-well-structured.
\end{itemize}
\end{theorem}
\Proof

\NI
$(\Ra)$
If for some node $i$ we have $top \not\in p(i)$,
then $i$ cannot adopt product $top$ and
$[top]$ is not reachable.

To establish the second condition consider a reduction
sequence
\[
p_{1} \myra p_{2} \myra \LL \myra p_{m}
\]
starting in $p$ and such that $p_{m} = [top]$.

Assign now to each node $i$ the minimal $k$ such that $p_{k+1}(i) =
\C{top}$.
We claim that this definition of the $\mathtt{level}$ function shows that $G_{p, top}$ is
($\theta,top$)-well-structured.
To see this, consider a node $i$.
\II

\NI
\emph{Case 1.} $\mathtt{level}(i) = 0$.

Then $p(i) = \C{top}$, and we have $N(i) = \ES$ in $G_{p, top}$.
Hence we do not need to argue about node $i$, since we only need to ensure (\ref{equ:theta}) for nodes with $N(i) \neq \ES$.
\II

\NI
\emph{Case 2.} $\mathtt{level}(i) > 0$.

Suppose that $N(i) \neq \ES$
and that $\mathtt{level}(i) = k$. By the
definition of the reduction $\myra$ 
the adoption condition $A(top,i)$ holds in $p_k$, i.e., 
\[
\sum_{j \in N(i)\mid p_k(j) = \{top\} } w_{ji} \geq  \theta(i,top).
\]


But for each $j \in N(i)$ such that $p_k(j) = \{top\}$ we have by
definition $\mathtt{level}(j) < \mathtt{level}(i)$.  So (\ref{equ:theta}) holds.
\II

\NI
$(\La)$
Consider a certificate function $\mathtt{level}$ showing that $G_{p, top}$ is ($\theta,top$)-well-structured.
Without loss of generality we can assume that the nodes in  $G_{p, top}$ such that $N(i) = \ES$
are exactly the nodes of level $0$.
We construct by induction on the level $m$ a reduction sequence
$p \tra p''$,
such that for all nodes $i$ we have $top \in p''(i)$ and for all nodes $i$
of level $\leq m$ we have $p''(i) = \C{top}$.

Consider level $0$.  By definition of $G_{p, top}$, a node $i$ is of level $0$ iff it has no
neighbours in $G$ or $p(i) = \C{top}$.  In the former case, by the
first condition, $top \in p(i)$. So $p \tra p''$, where the function
$p''$ is defined by
\[
        p''(i) :=
        \left\{
        \begin{array}{l@{\extracolsep{3mm}}l}
        \C{top}   & \mathrm{if}\  \mathtt{level}(i) = 0 \\
        p(i)      & \mathrm{otherwise}
        \end{array}
        \right.
\]

This establishes the induction basis. Suppose now that the claim holds for some level $m$, yielding the reduction sequence $p \tra
p'$.
Consider the nodes of level $m+1$.
For each such node $i$ we have  $top \in p'(i)$, $N(i) \neq \ES$ and
\[
\sum_{j \in N(i) \mid \mathtt{level}(j) < \mathtt{level}(i)} w_{ji} \geq \theta(i,top).
\]

By the definition of $G_{p, top}$
the sets of neighbours of $i$ in $G$ and $G_{p, top}$ are the same.
By the induction hypothesis
for all nodes $j$ such that $\mathtt{level}(j) < \mathtt{level}(i)$ we have $p'(j) = \{top\}$.
Hence, either such a node $i$ adopted product $top$ in $p'$ or
can adopt product $top$ in $p'$.

Thus, $p' \tra p''$, where
the function
$p''$ is defined by
\[
        p''(i) :=
        \left\{
        \begin{array}{l@{\extracolsep{3mm}}l}
        \C{top}   & \mathrm{if}\  \mathtt{level}(i) = m+1 \\
        p'(i)      & \mathrm{otherwise}
        \end{array}
        \right.
\]

Consequently $p \tra p''$, which establishes the induction step. By induction we conclude $p \tra [top]$.
\HB

\VV
Next we show that testing if a graph is ($\theta,t$)-well-structured can be
efficiently solved.

\begin{theorem}
\label{thm:theta}
Given a weighted directed graph $G$, a threshold function $\theta$ and a product $t$, we can decide whether
$G$ is ($\theta,t$)-well-structured in time $O(n^2)$.
\end{theorem}
\Proof
We claim that the following simple algorithm achieves this:

\begin{itemize}
\item Given a weighted directed graph $G$, first assign level $0$ to
  all nodes with $N(i) = \emptyset$. If no such node exists, output
  that the graph is not ($\theta,t$)-well-structured.

\item Inductively, at step $i$, assign level $i$ to each node for
  which condition (\ref{equ:theta}) from Definition~\ref{def:q} is
  satisfied when considering only its neighbours that have been
  assigned levels $0,\LL,i-1$.

\item If by iterating this all nodes are assigned a level, then output
  that the graph is ($\theta,t$)-well-structured. Otherwise,
output that $G$ is not ($\theta,t$)-well-structured.
\end{itemize}

The above algorithm can be implemented in time $O(n^2 + |E|) =
O(n^2)$.  We can first create the adjacency list
representation so that for each node we have a list with its outgoing
edges.  
Given this representation, we can implement the steps of the algorithm
in $O(|E|)$ time. The idea is that each edge of
the graph is processed only once and only a constant number of
operations is needed for every edge. Indeed, one can keep a counter
for every node that sums up the weight from nodes that have already
been assigned a level. For every node that was assigned a level at the
previous round, one can go through its outgoing edges and update the
corresponding counters accordingly (only counters of nodes that have not yet been assigned a level are updated). 
Hence we can assign a level number
to any node whose counter has been updated at the current round and
has exceeded the threshold.

To prove the correctness of the algorithm, note that if the input
graph is not ($\theta,t$)-well-structured, then the algorithm will output
No, as otherwise, at termination it would have constructed a $\mathtt{level}$ function for a
non-($\theta,t$)-well-structured graph. Hence it remains to prove that if
a graph is ($\theta,t$)-well-structured, the algorithm will output Yes.

Suppose a graph $G$ is ($\theta,t$)-well-structured.
We will use a certificate function, $l_G$, in which all nodes are assigned the minimum
possible level. For each node $i$, let $l^i$ be a certificate function where node $i$ has the minimum possible level.
Then define $l_G(i) := \min_j l^j(i) =l^i(i)$. 

First note that $l_G$ is a certificate function because a minimum of certificate functions is also
a certificate for $G$. By the definition of $l_G$, the level of each node $i$
cannot be lowered below $l_G(i)$, i.e., for all nodes $i$

\begin{equation}
\label{equ:geq}
l_G(i) = \min \{k : \sum_{j \in N(i) \mid l_G(j) < k} w_{ji} \geq \theta(i,t)\}.
\end{equation}

We now prove that every node is assigned a level by the algorithm and
in particular that $l_G$ is the function $\mathtt{level}$ generated by
the algorithm, hence the algorithm outputs Yes. For level $0$, note
that by the minimality of $l_G$ and since $\theta(i,t) > 0$ for every
$i$, the only nodes for which $l_G$ assigns $0$ are all nodes $i$ such
that $N(i) = \emptyset$. But these are precisely the nodes that are
assigned level $0$ by the algorithm as well.

Suppose by induction that $l_G$ and $\mathtt{level}$ coincide on all nodes
considered by the algorithm in steps $1, \LL, k-1$, where $k$ is a
level used by $l_G$. Then by the construction of the algorithm and by
(\ref{equ:geq}), the algorithm assigns level $k$ to {\em all} nodes
$i$ such that $l_G(i) = k$. Moreover, since $k$ is used by $l_G$, some
new nodes are assigned a level at step $k$.

Hence, $l_G$ and $\mathtt{level}$ coincide. Consequently the algorithm assigns a level to all nodes and hence outputs Yes.
\HB



\begin{note}
  The above algorithm can run in time $O(|E|)$ when
  we are given directly the representation of the graph in terms of
  adjacency lists of outgoing edges for each node instead of the adjacency matrix.
\end{note}

Finally, we end this section by observing that the algorithmic question of
determining whether a network $[top]$ is reachable can be solved efficiently.

\begin{theorem} \label{thm:res1-alg}
Assume a network $(G,P,p,\theta)$ and a product $top \in P$.
There is an algorithm running in time  $O(n^2)$ that determines whether
the network $(G,P,[top],\theta)$ is reachable.
\end{theorem}
\Proof The proof follows either by using Theorem \ref{thm:res1} and Theorem \ref{thm:theta} for $G_{p,top}$ or by simply start performing adoptions only of product $top$ until no further reduction is possible.
\HB

\section{Unavoidable outcomes}
\label{sec:unavoidable}

Next, we focus on the notion of unavoidable outcomes.
We establish the following characterization.
\begin{theorem} \label{thm:res2}
Assume a network $(G,P,p,\theta)$ and a product $top \in P$.
A network $(G,P,[top],\theta)$ is unavoidable iff
\begin{itemize}
\item for all $i$, if $N(i) = \ES$, then $p(i) = \{top\}$,

\item for all $i$, $top \in p(i)$,

\item $G_{p,top}$ is ($\theta,top$)-well-structured.
\end{itemize}
\end{theorem}

To prove this, we shall need first a few lemmas.

\begin{lemma} \label{lem:t}
Suppose that $p \tra p'$ and for some node
$i$ we have $p'(i) = \{t\}$.
Then for some node $j$ such that $N(j) = \ES$ or $p(j)$ is a singleton,
we have $t \in p(j)$.
\end{lemma}
Intuitively, this means that for each product eventually adopted by some node, the diffusion of the product must have started by some 
node $j$ (possibly different) such that either $j$ had already adopted it or the product was a possible choice for $j$ in the beginning of the sequence and $j$ has no neighbors.
\II

\NI
\Proof
Let $p  \tra p'$ be of the form
\[
p_{1} \myra p_{2} \myra \LL \myra p_{m}.
\]
Let $l'$ be the smallest index $l$ such that for some node $j$ we have $p_{l}(j) = \{t\}$.
If $l' = 1$, then $p(j) = \{t\}$ and we are done. If $l' > 1$
then by the choice of $l'$ we have $N(j) = \ES$, as otherwise for some node $k$ we would have
$p_{l'-1}(k) = \{t\}$ (recall that $\theta(i, t)\neq 0$ for all $i, t$ and that for nodes that have a non-empty neighborhood, the only way to adopt a product is through their neighbors). Moreover, $t \in p_{l'-1}(j)$ implies  $t \in p_1(j)$, that is $t \in p(j)$.
\HB

\begin{lemma} \label{lem:top}
Assume a network $(G,P,p,\theta)$ and a product $top \in P$.
Suppose that
\begin{itemize}
\item for all $i$, if $N(i) = \ES$ or $p(i)$ is a singleton, then $p(i) = \{top\}$.
\end{itemize}
Then $(G,P,p,\theta)$ admits a unique outcome.
\end{lemma}
Intuitively, this means that if initially only one product can be adopted, then
a unique outcome of the network exists.
\II

\NI
\Proof
Consider two maximal sequences of reductions $p \tra p'$ and $p
\tra p''$.
Let $p  \tra p'$ be of the form
\[
p_{1} \myra p_{2} \myra \LL \myra p_{m}.
\]
We prove by induction on $k$ that for all nodes $i$ and products $t$ if
$p_k(i) = \{t\}$, then $p''(i) = \{t\}$.
If $p(i) = \{t\}$, then also $p''(i) = \{t\}$.
This takes care of the induction basis.

Assume the claim holds for some $k$ and suppose
$p_{k+1}(i) = \{t\}$. If $p_{k}(i) = \{t\}$, then
by the induction hypothesis $p''(i) = \{t\}$.
Otherwise by the definition of the $\myra$ relation
$t \in p_{k}(i)$, $N(i) \neq \ES$ and $A(t,i)$ holds in $p_k$.

By the assumption and Lemma \ref{lem:t} $t = top$.
By the induction hypothesis $A(t,i)$ holds in $p''$.

Moreover, $p_{k+1}(i) = \C{top}$ implies $top \in p(i)$.
By the assumption and Lemma \ref{lem:t} for no $t' \neq top$ we have
$p''(i) = \C{t'}$.
Hence any maximal sequence of reductions from $p''$ will lead to a network $p^{\#}$, with $p^{\#}(i) = \C{top}$. But
$p \tra p''$ is a maximal sequence of reductions, so
$p^{\#} = p''$ and consequently $p''(i) = \C{top}$.

We conclude by induction that for all nodes $i$ if
$p'(i) = \{t\}$, then $p''(i) = \{t\}$.
By symmetry the reverse implication holds.
By the definition of the $\myra$ relation this implies
that $p' = p''$.
\HB
\VV

\noindent {\bf Proof of Theorem~\ref{thm:res2}}

\NI
$(\Ra)$
If $[top]$ is unavoidable, then it is reachable from $p$, hence, thanks to Theorem \ref{thm:res1} we only need to establish the first condition.
But if for some
node $i$ such that $N(i) = \ES$ we have $p(i) \neq \{top\}$,
then $i$ can adopt a different product than $top$ and
$[top]$ cannot be unavoidable.
\II

\NI
$(\La)$
By Theorem \ref{thm:res1} $[top]$ is reachable, so we only need to show that it is a unique outcome.
But this is guaranteed by Lemma \ref{lem:top}.
\HB
\VV

In analogy to Theorem~\ref{thm:res1-alg}, we also have the following simple fact.
\begin{theorem} \label{thm:res2-alg}
Assume a network $(G,P,p,\theta)$ and a product $top \in P$.
There is an algorithm, running in time $O(n^2)$, that determines whether
the network $(G,P,[top],\theta)$ is unavoidable.
\end{theorem}



\section{Unique outcomes}
\label{sec:unique}

\begin{figure}[htbp]
\begin{center} \ \setlength{\epsfxsize}{3.5cm}
\epsfbox{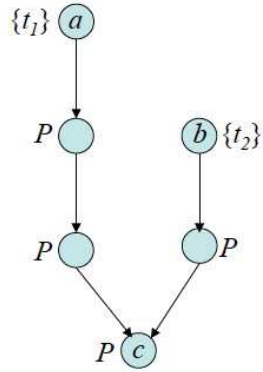}
\end{center}
\caption{An example of a network with a non-unique outcome, with $P = \{t_1, t_2\}$.} \label{fig:non-unique}
\end{figure}

Finally, we consider the question of when a network admits a unique
outcome.  In the definition of the reduction relation $\to$ between
networks we only stipulate that \emph{some} node adopts a product. In
several situations the order in which nodes adopt the products does
not matter. So this definition can introduce a `spurious'
nondeterminism in the sense that all maximal reduction sequences still
yield the same outcome. One way to eliminate this spurious
nondeterminism consists of employing a different form of reduction,
that we call below \emph{fast}, in which we stipulate that \emph{all}
nodes that can adopt a product do so.  A natural conjecture is then
that a network does not admit a unique outcome iff such a modified form of
reduction eventually yields a network in which a node can adopt more
than one product. However, such a conjecture is false.  As an example
consider the network in Figure \ref{fig:non-unique}.

In this network such a fast reduction eventually yields a network in which node $c$ adopts
product $t_2$. However, there is also another reduction sequence which results
in node $c$ adopting product $t_1$.
To rectify this conjecture we introduce the following concepts.

\begin{definition}
Given networks $p, p'$ based on the same graph we say that
\begin{itemize}
\item the reduction $p \myra p'$ is \bfe{fast} if for each node $i$, if
  $i$ can adopt a product in $p$ then $i$ adopted a product in $p'$. Intuitively, $p \myra p'$ is then a `maximal' one-step reduction of $p$,

\item
node $i$  \bfe{can switch in $p'$ given $p$} if $i$ adopted $t$ in $p'$ and for some $t' \neq t$
\[
\mbox{$t' \in p(i) \A \sum_{j \in N(i) \mid p'(j) = \{t'\}} w_{ji} \geq  \theta(i,t')$,}
\]

\item $p'$ is \bfe{ambivalent given $p$} if it contains a node that
  either can adopt more than one product or can switch in $p'$ given $p$.
\HB
\end{itemize}
\end{definition}

The definition of ambivalence above essentially describes the two
reasons that may cause a network not to admit a unique outcome.  Note
now that given the network in Figure \ref{fig:non-unique} after
performing two fast reductions we obtain a final network in which node
$c$ can switch to $t_1$. So the notion of switching allowed us to find
out that the initial network does not admit a unique outcome.  This is
the case in general. To formulate the general result we need one more
notion.

\begin{definition}
By the \bfe{contraction sequence} of a network we mean the unique
reduction sequence $p \tra p'$ such that
\begin{itemize}
\item each of its reduction steps is fast,

\item either $p \tra p'$ is maximal or $p'$ is the first network in the sequence $p \tra p'$
that is ambivalent given $p$.
\HB
\end{itemize}
\end{definition}

We now formulate a characterization of networks that admit
a unique outcome.

\begin{theorem} \label{thm:res3}
A network admits a unique outcome iff its contraction sequence ends in a
non-ambivalent network.
\end{theorem}

Comparing this characterization to the original false conjecture we
see that it still employs the fast reduction but refers to a different
`stopping criterion' that also takes into account the possibility of
switching.

\Proof

\NI
$(\Ra)$
Suppose that a network $p$ admits a unique outcome and assume
by contradiction that the contraction sequence $\chi$ of $p$ ends in
an ambivalent network $p'$. If a node in $p'$ can adopt two
different products, then we get a contradiction. Otherwise a node $i'$
in $p'$ can switch from a product $t$ to a product $t' \neq t$.

Given a reduction sequence $\xi$ that starts in $p$ and a node $i$
that adopted in it a product $t$, but not initially (so not in $p$),
we define a modified reduction sequence in which this node can adopt a
product but did not adopt any. This is done so as to cancel
all adoptions that led $i'$ to adopt $t$.
To this end we set $p''(j) :=p(j)$ for
every node $j$ that adopted product $t$ and every network $p''$ from
$\xi$ and subsequently remove from the resulting sequence the
duplicate networks.

Since node $i'$ can switch from $t$ to $t'$, we have $\{t,t'\} \sse
p(i')$, so on $\chi$ node $i'$ did not adopt the product $t$
initially.  So the corresponding modification of $\chi$ w.r.t. node
$i'$ results in a reduction sequence that starts in $p$ and in which
node $i'$ can adopt product $t'$. So $p$ admits two outcomes which
yields a contradiction.
\II

\NI
$(\La)$
First, given a maximal reduction sequence $\xi := p \tra p'$ we
define its \bfe{fast run} inductively by its length as follows.  If $p
= p'$, then $p$ is the fast run of $p \tra p'$. Otherwise, $\xi = p
\myra p_1 \tra p'$ for some network $p_1$.  Define a social
network $p_2$ as follows:
 \[
         p_2(i) :=
         \left\{
         \begin{array}{l@{\extracolsep{3mm}}l}
         \{t\}   & \mbox{if $i$ can adopt $t$ in $p$ and $p'(i) = \{t\}$} \\
         p(i)      & \mathrm{otherwise}
         \end{array}
         \right.
 \]
We have then $p \myra p_2$ and $p_2 \tra p'$. We define then the fast run of
$p \tra p'$ as the concatenation of $p \myra p_2$ and the fast run of $p_2 \tra p'$.

Intuitively, a fast run of a maximal reduction sequence $p \tra p'$ yields the same
final result, $p'$, but by maximizing at each reduction step the number of nodes
that adopt a product.

Suppose now that the contraction sequence of a network $p$ ends in a
non-ambivalent network and assume by contradiction that $p$
admits two outcomes. So two sequences of reductions $\xi$ and $\xi'$
exist that both start in $p$, are maximal, and their final elements
differ.

Let $fr(\xi)$ and $fr(\xi')$ be the respective fast runs of $\xi$ and
$\xi'$.  By assumption at least one of these two fast runs, say
$fr(\xi)$, differs from the contraction sequence $\chi$ of $p$. Let
$p'_1$ be the first network in the sequence $\chi$ in which a
difference with $fr(\xi)$ arises.

By assumption $p'_1$ is non-ambivalent, so some fast reduction $p'_1
\myra p'$ is part of $\chi$ and a reduction $p'_1 \myra p''$ with $p'
\neq p''$ is part of $fr(\xi)$.  Since $p'_1 \myra p'$ is a fast
reduction and $fr(\xi)$ is a fast run, the difference between $p'$ and
$p''$ arises due to the fact that some node $i$ adopted in $p'$
one product and in $p''$ a different product.  But this means that
$p'_1$ is ambivalent, which is a contradiction.
\HB
\VV

Theorem~\ref{thm:res3} yields a simple algorithm for testing whether a network has a unique outcome.
\begin{theorem} \label{thm:res3-alg}
  There exists a polynomial time algorithm, running in time $O(n^2 + n|P|)$, that determines whether a
  network admits a unique outcome.
\end{theorem}

For all practical purposes we have $|P| \ll n$, so the running time
is in practice $O(n^2)$. 
\II

\NI
\textbf{Proof of Theorem \ref{thm:res3-alg}}.
By Theorem \ref{thm:res3} it suffices to determine whether the
contraction sequence of a network $p$ ends in a non-ambivalent social
network.  This can be tested using the algorithm presented in
Figure~\ref{alg}.  The algorithm keeps performing fast reductions
until we realize that either a node can adopt two different products
or can switch from one product to another. If none of these happens
then we can safely conclude given Theorem~\ref{thm:res3} that the
network has a unique outcome.

Given a network $(G, P, p, \theta)$, the algorithm uses for
each node $j$ and each product $t\in p(j)$ a counter $S_{j,t}$. The
counter measures the accumulated weight from incoming edges that have
already adopted a product $t$.

Regarding the complexity of the algorithm, the initial part of
producing the required representation in Line 1 may take time up to
$O(n^2)$ if we are given the matrix representation or any other of the
standard ways of representing a graph. The initialization of the
counters $S_{j,t}$ requires in the worst case $O(n|P|)$. As for the
remaining part, the variable $L$ maintains the set of nodes that
adopted a product in the last round (Lines 11 and 28). Each edge
$(i,j)$ is examined exactly once, just after $i$ adopts a product.

The number of operations that we need to perform for every edge is
$O(1)$ because we only need to update the appropriate counter
$S_{j,t}$ and add $j$ to the list $R$ (Lines 16-17). Furthermore, we
also need to check for each such $j$ whether it can adopt more than
one product. This can also be done while we update each counter
$S_{j,t}$ by having another counter that increases by one for every
$S_{j,t}$ that exceeds the threshold $\theta(j,t)$. In total, we do not
need more than $O(1)$ operations per edge and therefore the total
running time is $O(n^2 + n|P| +|E|)$ = $O(n^2 + n|P|)$.  
\HB

\begin{figure}[htbp]
\hrule height0.8pt\vspace{5.8pt}
\begin{algorithmic}[1]

\STATE Produce the representation with a list of outgoing edges for each node;
\FOR{$i \in V$} 
\STATE Set $p(i)$ to be the initial list of products available to node $i$; 
\ENDFOR
\FOR{$j\in V, t\in p(j)$}
\STATE $S_{j,t} := 0$; // counts total weight to $j$ from nodes that adopted $t$
\ENDFOR
\IF{$\exists i\in V$ with $N(i) = \emptyset$ and $|p(i)|\geq 2$}
\RETURN "No unique outcome";
\ENDIF
\STATE $L := \{ i\in V: |p(i)| = 1\}$; // initialize $L$ to a list of nodes that already have adopted a product
\STATE \textbf{if} {$L = \emptyset$} \textbf{return} "Unique outcome"; \textbf{endif}
\WHILE{$L\neq \emptyset$}
\STATE $R := \emptyset$;
\FOR{$i\in L$ and $j$ such that $(i, j)\in E$}
\STATE \textbf{if} $i$ has adopted $t$ and $t\in p(j)$ \textbf{then} $S_{j,t} := S_{j,t} + w_{ij}; $\textbf{ end if}
\STATE $R := R\cup \{j\}$; // nodes we need to check for ambivalence
\ENDFOR

\FOR{$j\in R$}
\STATE Compute $|\{t: S_{j, t} \geq \theta(j,t)\}|$; // even for nodes that have already adopted a product
\STATE \textbf{if} {$|\{t: S_{j, t} \geq \theta(j,t)\}| \geq 2$} \textbf{return} "No unique outcome"; \textbf{endif}
\IF{$|\{t: S_{j, t} \geq \theta(j,t)\}| = 1$ and $j$ has not yet adopted $t$}
\STATE node $j$ adopts product $t$;
\ELSE
\STATE $R := R\setminus \{j\}$;  // node $j$ does not adopt any product
\ENDIF
\ENDFOR
\STATE $L := R$; // put in $L$ all nodes that adopted a product in last round
\ENDWHILE
\RETURN "Unique outcome" // No further reduction is possible

\end{algorithmic}
\vspace{5pt}\hrule height 0.8pt
\caption{Pseudocode for the algorithm of Theorem~\ref{thm:res3-alg}} \label{alg}
\end{figure}
\VV

It would be interesting to find a structural characterization of networks that admit a unique outcome,
as Theorem \ref{thm:res3} only provides such a characterization in terms of the contraction sequences.
At this stage we only have the following result.

\begin{corollary}
\label{cor:unique}
Assume a network $(G,P,p,\theta)$ such that
for all nodes $i$ and products $t$ we have $\theta(i,t) > \frac{1}{2}$,

Then $(G, P, p, \theta)$  admits a unique outcome iff
for all $i$, $N(i) = \ES$ implies that $p(i)$ is a singleton.
\end{corollary}

\Proof
The ($\Ra)$ implication is obvious. For the ($\La)$ implication 
it suffices to note that if $p \tra p'$, then $p'$ is not ambivalent given $p$.
So the result is a direct consequence of Theorem \ref{thm:res3}.
\HB
\VV




This corollary can be strengthened by assuming that the network is
such that if for some product $t$ we have
$\theta(i,t) \leq \frac{1}{2}$, then $|N(i)| < 2$ or $|p(i)|
= 1$. The reason is that the nodes for which $|N(i)| < 2$ or $|p(i)| =
1$ cannot introduce an ambivalence.

When for some node $i$ and product $t$, $\theta(i,t) \leq \frac{1}{2}$ holds and neither
$|N(i)| < 2$ nor $|p(i)| = 1$, the equitable
network still may admit a unique outcome but it does not have
to.  For instance the second network in Figure \ref{fig:soc}
admits a unique outcome for the last three alternatives (explained in Example~\ref{ex:nets}),
while for the first two it does not.

Finally, we note that for the class of networks of Corollary~\ref{cor:unique} we have a 
simpler algorithm, removing the dependency on $|P|$. 
\begin{theorem} \label{thm:unique}
  There exists an algorithm, running in time $O(n^2)$, that determines whether a
  network, such that for all nodes $i$ and products $t$
we have $\theta(i,t) > \frac{1}{2}$,
 admits a unique outcome.
\end{theorem}

\Proof
By Corollary~\ref{cor:unique}.
\HB



\section{Product adoption}
\label{sec:adoption-analysis}

In this section we study a number of questions concerning adoption of the products
by the nodes of a given network, focusing on complexity matters.
Recall that given an initial network $p$, a final network is one that
has been obtained from $p$ by a maximal sequence of reductions.
We first clarify the complexity of the following problem.
\II

\NI {\bf FINAL:} Given an initial network determine whether
a final network exists in which every node adopted some product.

Note that in the definition of the problem, we do not insist that all nodes adopt the same product, but only that every node has eventually chosen a product.

\begin{theorem}
\label{thm:final}
\textbf{FINAL} is NP-complete, even for 2 products and product independent thresholds.
\end{theorem}

\Proof
First we prove that \textbf{FINAL} is in NP. Given an initial network, the
certificate can consist of a final network in which every node adopted
some product along with the series of reductions that led to this
final network (there can be at most $O(n)$ such reductions). One can
then check in polynomial time that this is a valid final network,
given the initial network, and that indeed all nodes have adopted a
product.

For NP-hardness, we give a reduction from the NP-complete PARTITION
problem, which is: given $n$ positive rational numbers
$(a_1,\LL,a_n)$, is there a set $S$ such that $\sum_{i\in S} a_i =
\sum_{i\not\in S} a_i$?  Consider an instance $I$ of PARTITION.
Without loss of generality, suppose we have normalized the numbers so
that $\sum_{i=1}^n a_i = \frac12$. Hence the question is to decide
whether there is a set $S$ such that $\sum_{i\in S} a_i =
\sum_{i\not\in S} a_i = \frac{1}{4}$.

\begin{figure}[htbp]
\begin{center}  \ \setlength{\epsfxsize}{5cm}
\epsfbox{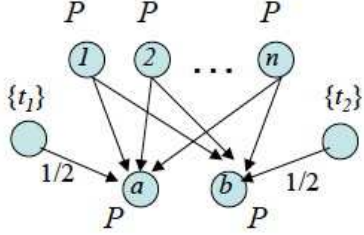}
\end{center}
\caption{Social network related to the \textbf{FINAL} problem, with $P = \{t_1, t_2\}$.}
\label{fig:final}
\end{figure}

We build an instance of our problem with two products, namely $P =
\{t_1, t_2\}$, and with the network shown in
Figure~\ref{fig:final}. 
The threshold function does not depend on the product argument (that
is omitted) and is given by: $\theta(a) = \theta(b) = \frac{3}{4}$.  
Finally, for each node $i\in\{1,\LL,n\}$,
we set $w_{i a} = w_{i b} = a_i$. The weights of the other two edges
are $\frac12$.

Suppose there is a solution $S$ to $I$. Then we can have the nodes
corresponding to the set $S$ adopt $t_1$ and the remaining nodes from
$\{1,\LL,n\}$ adopt $t_2$. By the choice of the weights $w_{i a}$
and $w_{i b}$ this implies that node $a$ can adopt $t_1$
and node $b$ can adopt $t_2$.  Hence a final network exists in which
all nodes adopted a product.

For the reverse direction, suppose that a final network exists in
which all nodes adopted a product. Then node $a$ adopted product $t_1$
and node $b$ adopted product $t_2$, as it is not possible for node $a$
to adopt $t_2$ and for node $b$ to adopt $t_1$. 
Let $S$ be the set of nodes
$i \in \{1, \LL, n\}$ that adopted product $t_1$. Then by the choice
of the weights $w_{i a}$ and $w_{i b}$ and the thresholds of the nodes
$a$ and $b$, it holds that both $\sum_{i\in S} a_i \geq \frac{1}{4}$
and $\sum_{i\not\in S} a_i \geq \frac{1}{4}$.  But since $\sum_{i=1}^n
a_i = \frac12$, this implies that $\sum_{i\in S} a_i = $ $\sum_{i\not\in S}
a_i = \frac{1}{4}$, i.e., there is a solution to the instance $I$ of
the PARTITION problem.
\HB
\VV

We now move on to a different class of problems, motivated by the results of Section \ref{sec:unique}, which reveal that many networks will not
admit a unique outcome. Therefore the following questions concerning
product adoption by a given node are of natural interest for such networks.
\II

\NI {\bf ADOPTION 1:} (unavoidable adoption of some product)  \\
Determine whether a given node has to adopt some product in all final networks.

\NI {\bf ADOPTION 2:} (unavoidable adoption of a given product) \\
Determine whether a given node has to adopt a given product in all final networks.

\NI {\bf ADOPTION 3:} (possible adoption of some product) \\
Determine whether a given node adopted some product in some final network.

\NI {\bf ADOPTION 4:} (possible adoption of a given product) \\
Determine whether a given node adopted a given product in some final network.

Below we resolve the complexity of all these problems.
\begin{theorem}
\label{thm:adoption-problems}
The complexity of the above problems is as follows:
\begin{enumerate}[(i)]
\item \textbf{ADOPTION 1} is co-NP-complete, even for 2 products and product independent thresholds.

\item \textbf{ADOPTION 2} for 2 products can be solved in $O(n^2)$ time.

\item \textbf{ADOPTION 2} is co-NP-complete for at least $3$ products, even with product independent thresholds.

\item \textbf{ADOPTION 3} can be solved in $O(n^2|P|)$ time.

\item \textbf{ADOPTION 4} can be solved in $O(n^2)$ time.
\end{enumerate}
\end{theorem}

\Proof

\NI $(i)$
It suffices to prove NP-completeness of the complementary problem,
which is: given an initial network determine if there is a final
network in which a given node does not adopt any product.  The
argument for the membership in NP is very similar to the membership
proof in Theorem \ref{thm:final}.

\begin{figure}[htbp]
\begin{center}  \ \setlength{\epsfxsize}{3.5cm}
\epsfbox{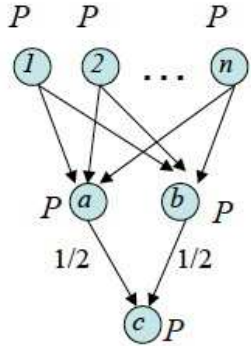}
\end{center}
\caption{Social network related to the \textbf{ADOPTION 1} problem. Here $P = \{t_1, t_2\}$.}
\label{fig:adoption}
\end{figure}

To prove NP-hardness, we use again a reduction from the PARTITION
problem but with a different normalization for the PARTITION instance.
In particular, we assume an instance $I$ with the numbers
$a_1,...,a_n$ satisfying $\sum_{i=1}^n a_i = 1$.  Hence the question
is to decide whether there is a set $S$ such that $\sum_{i\in S} a_i =
\sum_{i\not\in S} a_i = \frac{1}{2}$.  
We also use a slightly
different network than the one depicted in Figure \ref{fig:final}.
Given the instance $I$ we use the network presented in Figure
\ref{fig:adoption}.  This network depicts an instance of our problem
with $P = \{t_1, t_2\}$ and with node
$c$ as the "designated" node. The threshold function does not depend
on the product argument (that is omitted) and is given by: $\theta(a)
= \theta(b) = \frac12$, and $\theta(c) = 1$. Finally, as in the
proof of Theorem \ref{thm:final}, for each node $i\in\{1,\LL,n\}$, we
set $w_{i a} = w_{i b} = a_i$. We also use the weights $w_{a c} = w_{b
  c} = \frac12$.

Suppose now that there is a solution $S$ to the PARTITION instance.
Then we can have the nodes corresponding to the set $S$ adopt $t_1$
and the remaining nodes from $\{1,\LL,n\}$ adopt $t_2$. By the choice
of the weights $w_{i a}$ and $w_{i b}$ and the thresholds of $a$ and
$b$, this implies that node $a$ can adopt $t_1$ and node $b$ can adopt
$t_2$. In that case node $c$ cannot adopt any product. Thus a final
network exists in which node $c$ does not adopt any product.

Suppose now that in a final network node $c$ did not adopt any
product.  Then it cannot be the case that nodes $a$ and $b$ adopted
the same product since then node $c$ would have adopted it as well.
Note also that in all final networks nodes $a$ and $b$ have adopted
some product. Suppose without loss of generality that node $a$ adopted
$t_1$ and node $b$ adopted $t_2$.  Let $S$ be the set of nodes $i \in
\{1, \LL, n\}$ that adopted $t_1$.  Then the nodes $i \in \{1, \LL,
n\} \setminus S$ adopted $t_2$.  By the choice of the weights we have
both $\sum_{i\in S} a_i \geq \frac{1}{2}$ and $\sum_{i\not\in S} a_i
\geq \frac{1}{2}$.  But since $\sum_{i=1}^n a_i = 1$, this implies
that $\sum_{i\in S} a_i = \sum_{i\not\in S} a_i = \frac{1}{2}$, i.e.,
there is a solution to the instance $I$ of the PARTITION problem.
\vspace{.1in}

\NI $(ii)$ The algorithm resembles the one used in the proof of
Theorem~\ref{thm:res3-alg}. Let $P = \{t_1, t_2\}$, and suppose
the given product is $t_1$. We use the following observation.
To determine whether a given node has to adopt $t_1$ in all final networks,
it suffices to check this for the worst possible
final network with respect to adoption of $t_1$. 
So we first perform fast reductions only for product $t_2$.
Once no further adoption of
$t_2$ is possible, we perform all possible adoptions of $t_1$ so as to
reach a final network. If in this final network, the given node has
not adopted $t_1$, the answer to \textbf{ADOPTION 2} is No and
otherwise the answer is Yes.
\vspace{.1in}

\NI $(iii)$ We provide a reduction from PARTITION but with a slightly more involved network than in the proof  of $(i)$. 
Note that again, it suffices to prove NP-completeness of the complementary problem,
which is: given an initial network determine if there is a final
network in which a given node does not adopt the given product.  
The argument for the membership in NP is straighforward.

\begin{figure}[htbp]
\begin{center}  \ \setlength{\epsfxsize}{4.5cm}
\epsfbox{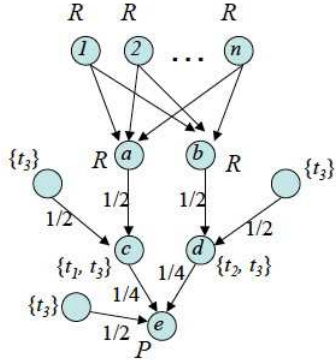}
\end{center}
\caption{Social network related to the \textbf{ADOPTION 2} problem. We fix $R = \{t_1, t_2\}$.}
\label{fig:adoption2}
\end{figure}

To prove NP-hardness, we start again with a PARTITION instance
$I$ with the numbers $a_1,...,a_n$ satisfying $\sum_{i=1}^n a_i = 1$. From this we construct the network shown in Figure \ref{fig:adoption2}.
This network has $3$ products, $P = \{t_1, t_2, t_3\}$, the designated node is $e$ and the designated product is $t_3$. The weights in the first layer of the graph are as in $(i)$. The rest of the weights are shown in the figure. The product independent threshold function is given by $\theta(a) = \theta(b) = \theta(c) = \theta(d) = \frac{1}{2}$, $\theta(e) = 1/2+\epsilon$, for some $\epsilon>0$.

Suppose there is a solution $S$ to $I$. Then we can have the nodes
corresponding to the set $S$ adopt $t_1$ and the remaining nodes from
$\{1,\LL,n\}$ adopt $t_2$. Then node $a$ can adopt $t_1$ and
node $b$ can adopt $t_2$. Subsequently, node $c$ can adopt $t_1$ and node
$d$ can adopt $t_2$. This yields a final network in which node $e$ does not adopt
product $t_3$.

Conversely, suppose that in a final network node $e$ did not adopt
product $t_3$. Then neither node $c$ nor node $d$ adopted $t_3$.
Hence node $c$ adopted $t_1$ and node $d$ adopted $t_2$ and
consequently node $a$ adopted $t_1$ and node $b$ adopted $t_2$.  As in
the proof of part $(i)$ this implies that there is a solution to the
instance $I$ of the PARTITION problem.  
\vspace{.1in}

\NI $(v)$ 
The algorithm resembles the one used in the proof of
Theorem~\ref{thm:res3-alg}.  Given a product, say $t$, it suffices to
start with the nodes that have already adopted $t$, perform fast
reductions but only with respect to $t$ until no further adoption of
$t$ is possible, and check if the given node has adopted $t$.
\vspace{.1in} 

\NI $(iv)$ Run the algorithm used in $(v)$ for each product.
\HB
\VV

It is interesting to observe the separation between \textbf{ADOPTION 1} and \textbf{ADOPTION 2} for $|P|=2$. While for $|P|\geq 3$ both problems are co-NP-complete and the proofs are based on similar arguments, in the case that $|P|=2$, \textbf{ADOPTION 2} becomes efficiently solvable but \textbf{ADOPTION 1} remains co-NP-complete.

We conclude our study by the following two optimization problems.
Suppose that a given product $top$ is neither reachable by all nodes nor
unavoidable for all nodes.  We would like then to estimate what is the
worst and best-case scenario for the spread of this product. That is,
starting from a given initial network $p$, what is the minimum
(resp.~maximum) number of nodes that will adopt this product in a
final network.  Hence, the following two problems are of interest.
\II

\NI 
{\bf MIN-ADOPTION:} Given an initial network and a product $top$,
what is the minimum number of nodes that adopted $top$ in a final
network.

\NI 
{\bf MAX-ADOPTION:} Given an initial network and a product $top$, what
is the maximum number of nodes that adopted $top$ in a final network.



We show that these two problems are substantially different, the first
being essentially inapproximable, while the second being efficiently
solvable.

\begin{theorem}
\label{thm:min-max-adoption}
Suppose  $n$ is the number of nodes of a network.
\begin{enumerate}[(i)]
\item \textbf{MAX-ADOPTION} can be solved in $O(n^2)$ time.
\item \textbf{MIN-ADOPTION} for $2$ products can be solved in $O(n^2)$ time.
\item For at least $3$ products and even with product independent thresholds, it is NP-hard to approximate \textbf{MIN-ADOPTION} with an approximation ratio better than $\Omega(n)$.
\end{enumerate}
\end{theorem} 
\Proof

\NI $(i)$
The algorithm is analogous to the one used when analyzing the
\textbf{ADOPTION 4} problem in the proof of Theorem
\ref{thm:adoption-problems}. Given a product $t$, we start with
the nodes that have already adopted the product and perform fast
reductions but only with respect to $t$ until no further adoption of
$t$ is possible.  
\vspace{.1in}

\NI $(ii)$ Suppose $P = \{t_1, t_2\}$ and that $t_1$ is the designated product. 
We first solve the \textbf{MAX-ADOPTION} problem for product $t_2$ and
then perform any necessary adoptions of $t_1$ to reach a final network.
This yields a final network with the minimum number of adoptions for
product $t_1$, since we only perform adoptions of $t_1$ that are necessary to reach a final network after first flooding the network as much as possible with the competitor. Note that this cannot be extended to the case of more products because then the order in which the competitors of $t_1$ spread may play a crucial role for minimizing the adoptions of product $t_1$. Hence it is not easy to determine a priori the sequence of adoptions of the products competing $t_1$.
\vspace{.1in}

\begin{figure}[htbp]
\begin{center}  \ \setlength{\epsfxsize}{6cm}
\epsfbox{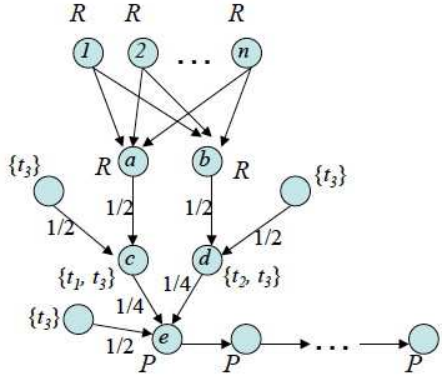}
\end{center}
\caption{The graph of the reduction with $P = \{t_1, t_2, t_3\}$ and $R = \{t_1, t_2\}$.}\label{fig:reduction-|P|=3}
\end{figure}

\NI $(iii)$
We again give a reduction from PARTITION, though the
appropriate network is now more involved. Consider an instance $I$ of
PARTITION problem, so $n$ positive rational numbers $(a_1,\LL,a_n)$
such that $\sum_{i =1}^{n} a_i = 1$.  We build an instance of our
problem with $3$ products, namely $P = \{t_1, t_2, t_3\}$, and with the
network shown in Figure~\ref{fig:reduction-|P|=3}. Note that this is derived by adding to the network of Figure \ref{fig:adoption2} 
a chain of M nodes starting from node $e$. We take $M$ to be $n^{O(1)}$ so
that the reduction is of polynomial time. The weight of each edge in the chain
is set to $1$.

We consider $t_3$ as the designated product. The threshold function does not depend on the product argument (that
is omitted) and is given by: $\theta(a) = \theta(b) = \theta(c) =
\theta(d) = \frac{1}{2}$, $\theta(e) = 1/2+\epsilon$, for some
$\epsilon>0$ and for the nodes to the right of node $e$ we can set the
thresholds to an arbitrary positive number in $(0,1]$.  Finally, for
each node $i\in\{1,\LL,n\}$, we set $w_{i a} = w_{i b} = a_i$. The
weights of the other edges can be seen in the figure.

We claim that if there exists a solution to the instance $I$, then a final
network exists with the number of nodes that adopted $t_3$ equal to
$3$, and otherwise in all final networks the number of nodes
that adopted $t_3$ equals $M+5$. This claim directly yields the
desired result, since $M = \Omega(|V|)$.

Suppose there is a solution $S$ to $I$.  As in the proof of
Theorem~\ref{thm:adoption-problems}$(iii)$ it follows that there
exists a final network in which node $e$ did not adopt product $t_3$.
Hence a final network exists in which only 3 nodes adopted $t_3$.

For the reverse direction, suppose there is no solution to the
PARTITION problem. This means that no matter how we partition the
nodes $\{1,\LL,n\}$, into two sets $S, S'$, it will always be that for
one of them, say $S$, we have $\sum_{i\in S} a_i > \frac{1}{2}$,
whereas for the other we have $\sum_{i\in S'} a_i < \frac{1}{2}$.
Thus in each final network, no matter which nodes from $\{1,\LL,n\}$
adopted $t_1$ or $t_2$, the nodes $a$ and $b$ adopted the same product.
Suppose for example that nodes $a$ and $b$ both adopted $t_1$ (the same
reasoning applies if they both adopted $t_2$). This in turn implies that
node $c$ adopted $t_1$ and node $d$ did not adopt $t_2$. Thus, 
the only possibility for node $d$ is to adopt $t_3$. But then the only choice for node
$e$ is to adopt $t_3$ and this propagates along the chain starting from node
$e$. This completes the proof. 
\HB 
\VV

\section{Structural results}
\label{sec:structural}

In~\cite{AM11} we used a slightly more restricted model of a social network
in that the threshold functions were product independent.
We clarify here the relation between these two models by 
presenting two transformations of the social networks
considered here, to social networks with product independent
threshold functions and by explaining in which sense they are related.

The first transformation takes as input an arbitrary social network $\snet
:= (G,P,p,\theta)$ and produces an equitable social network with
threshold functions that do not depend on the product argument.

First we add a new product $t_0$ to $P$. Then for each node $i$ such that $N(i) \neq \ES$
and $|p(i)| \geq 2$ we remove the edges $j \to i$ for each node $j \in N(i)$
and perform the following steps for each product $t \in p(i)$ and
each minimal subset $S$ of $N(i)$ such that
$
\sum_{j \in S \mid p(j) = \{t\} } w_{ji} \geq \theta(i,t):
$

\begin{itemize}
\item add a new node $a_{S,t,i}$,

\item put $p(a_{S,t,i}) := \{t, t_0\}$ and $\theta(a_{S, t,i}) := 1$,

\item add the edges $j \to a_{S,t,i}$ for each node $j \in S$,

\item add the edge $a_{S,t,i} \to i$,

\item put $\theta(i) := \frac{1}{2^{|N(i)|} |p(i)|}$.
\end{itemize}

Call the resulting equitable network $\snet'$. Intuitively, in this new network, a node $i$ can adopt a product $t$ when $t$ is adopted by any one of the nodes $a_{S,t,i}$, described above, which represent minimal subsets that can cause adoption. The new product independent threshold of node $i$ is set in a way that any such set $S$ can make node $i$ adopt a product. Finally, we also need to add the product $t_0$ to $p(a_{S,t,i})$ so that this node does not automatically adopt $t$.  

The following result relates the networks $\snet$ and $\snet'$.  We
say here that $p'_0$ is an \bfe{extension} of a network $p_0$ if it
can be obtained from $p_0$ by repeatedly replacing an edge $j \to i$
by a set of edges $\{j \to k, k \to i \mid k \in S_{i,j}\}$, where
$S_{i,j}$ is a set of new nodes, each with an appropriate set of
products and a threshold function.  $p'_0$ is then interpreted as an
equitable network.  We also say then that $p_0$ is a
\bfe{restriction} of $p'_0$.

\begin{theorem} \label{thm:arbitrary}
Consider the networks $\snet$ and $\snet'$. 

\begin{enumerate}[(i)]
\item If $\snet \tra p_0$ for a final network $p_0$ given $\snet$,
then for an extension $p'_0$ of $p_0$ we have $\snet' \tra p'_0$, where 
$p'_0$ is a final network given $\snet'$.

\item If $\snet' \tra p'_0$ for a final network $p'_0$ given $\snet'$,
then for a restriction $p_0$ of $p'_0$ we have $\snet \tra p_0$, where
$p_0$ is a final network given $\snet$.
\end{enumerate}

\end{theorem}

\Proof 

\NI
$(i)$ Let $i$ be the first node belonging to $\snet$ and such that
$N(i) \neq \ES$, $|p(i)| \geq 2$ and $i$ adopted a product $t$ in the
reduction sequence $\snet \tra p_0$.  So for some subset $S$ of $N(i)$
we have 
$
\sum_{j \in S \mid p(j) = \{t\} } w_{ji} \geq \theta(i,t).
$
Choose a minimal subset $S$ with this property.
Then by the definition of the threshold functions
in the network $\snet'$ node $a_{S,t,i}$ can adopt product $t$ and subsequently 
node $i$ can adopt $t$, as well.
Repeating this procedure we obtain the desired extension $p'_0$ of $p_0$.
\II

\NI
$(ii)$ Let $i$ be the first node belonging to $\snet$ such that
$N(i) \neq \ES$, $|p(i)| \geq 2$ and $i$ adopted a product $t$ in the
reduction sequence $\snet' \tra p'_0$.  So product $t$ was first
adopted in $\snet'$ by some node $a_{S, t, i}$ and then by $i$. By the definition of the
thresholds functions in $\snet'$ node $i$ can adopt product $t$ in the network
$\snet$.  Repeating this procedure we obtain the desired restriction $p_0$ of $p'_0$.
\HB
\VV

A disadvantage of this transformation is that it yields an exponential
blow up in the number of nodes.  Indeed, $\snet'$ has in the worst
case $n + n 2^n |P|$ nodes, where $n$ is the number of nodes in $\snet$.

A smaller increase can be achieved by the second transformation that
takes as input an equitable network $\snet := (G,P,p,\theta)$.
First we add a new product $t_0$ to $P$. Then for each node $i$ such
that $N(i) \neq \ES$ and $|p(i)| \geq 2$ we remove the edges $j \to i$
for each node $j \in N(i)$ and perform the following steps for each
product $t \in p(i)$:

\begin{itemize}

\item add a new node $a_{t,i}$,

\item put $p(a_{t,i}) := \{t, t_0\}$ and $\theta(a_{t,i}) := \theta(i, t)$,

\item add the edges $j \to a_{t,i}$ for each node $j \in N(i)$,

\item add the edge $a_{t,i} \to i$,

\item put $\theta(i) := \frac{1}{|p(i)|}$.
\end{itemize}

Call the resulting network $\snet'$. Note that $\snet'$ has
$\leq n (|P|+1)$ nodes, where $n$ is the number of nodes in $\snet$. The intuition behind this construction is simply that, unlike the first transformation,
we do not need to argue separately about each minimal subset that can cause further adoptions.
The following result relates the networks $\snet$ and $\snet'$.

\begin{theorem} \label{thm:equitable}
Consider the equitable networks $\snet$ and $\snet'$. 

\begin{enumerate}[(i)]
\item If $\snet \tra p_0$ for a final network $p_0$ given $\snet$,
then for some extension $p'_0$ of $p_0$ we have $\snet' \tra p'_0$, where 
$p'_0$ is a final network given $\snet'$.

\item If $\snet' \tra p'_0$ for a final network $p'_0$ given $\snet'$,
then for a restriction $p_0$ of $p'_0$ we have $\snet \tra p_0$, where
$p_0$ is a final network given $\snet$.
\end{enumerate}

\end{theorem}

\Proof
The proof is analogous to that of Theorem \ref{thm:arbitrary} and omitted.
\HB

\section{Conclusions and future work}

We have introduced a diffusion model in the presence of multiple
competing products and studied some basic questions.  We have provided
characterizations of the underlying graph structure for determining
whether a product can spread or will necessarily spread to the whole
network, and of the networks that admit a unique outcome. We
also studied the complexity of various problems that are of interest
for networks that do not admit a unique outcome, such as the problems
of computing the minimum or maximum number of nodes that will adopt a
given product in a final network, or the problem of determining whether
a given node has to adopt some (resp.~a given) product in all final
networks.

Our model of a social network attempts to capture a realistic
situation in which customers can select out of a number of products
and are influenced in their selection by their acquaintances and
friends.  Having this in mind it is useful to reflect on some of our
results from the viewpoint of a company trying to sell a specific product.
First, Theorem~\ref{thm:min-max-adoption}$(iii)$ shows that the
precise assessment of the worst case scenario is computationally
difficult.  In contrast, the computation of the best case scenario
(item $(i)$) is easy. Next, Theorems \ref{thm:res1-alg} and
\ref{thm:res2-alg} show that it is computationally easy to check
whether an adoption of a single product is possible or unavoidable.

However, one should bear in mind that `computationally easy' refers to
the time complexity $O(n^2)$ that can be too high for huge networks.
Still such an analysis may be useful for specialized networks, for
instance those concerning choice of a supermarket or a primary school
in a neighbourhood.

We conclude by listing some further topics of recent and future work.

\paragraph{Game theoretic analysis}

A natural follow up to this work is a game theoretic analysis that
aims at predicting customer behaviour. Such an analysis for players
choosing between two products has been presented in~\cite{Mor00}. An
extension with the additional option of adopting both products has
also been considered in~\cite{IKMW07}
(e.g. choosing to have two operating systems in your PC, instead of
just one). 

Recently \cite{SA12}, and more fully \cite{SA13}, used the model introduced here to study
consequences of adopting products by the nodes forming a social
network. This led to a study of strategic games in which the nodes
decide which product to choose (or decide not to adopt any).  In
particular, deciding whether a game in this class has a pure Nash
equilibrium is NP-complete, while deciding whether a game has the
so-called finite improvement property is co-NP-hard.

Other game-theoretic approaches can also be considered based on our
model.  In particular, one can consider a strategic game between the
producers who decide to offer their products for free to some selected
nodes.  A limited case was studied in \cite{AFPT10} (see also
\cite{THS12}) in a simpler model in which the thresholds were all
equal to $1/|N(v)|$ for every node $v$ (i.e., one neighbor suffices to
infect you).  Some recent follow up works on richer models include
\cite{GK12} and \cite{TAM12}.  An interesting direction here is to
obtain a better understanding of the structure of Nash equilibria
(pure or mixed), or identify conditions that guarantee the existence
of pure equilibria.

\paragraph{Introducing new products}
When a new product is introduced in a market, it is natural to assume
that this takes place when various customers have already adopted some
other product. The issue is then whether some nodes would switch to the new
product.  The present model does not allow us to study such a problem
since the input network for such an analysis is already a final
network and we stipulate that the choices of the nodes are final.
Allowing a switching by a node to a new
product can result in an `illegal' network, in which choices of some
nodes are not anymore justified and have to be reconsidered.

In contrast, in the framework of \cite{SA12} and \cite{SA13} such a study is possible,
since the input is simply a strategy profile that is an assignment of
products to nodes (with a special `no-choice' strategy allowed).
Recently we studied  in this setting, in \cite{AMS13}, the consequences of
introducing new products by means of improvement paths in the sense of
\cite{MS96}, the special case of which is the best-response dynamics.





\paragraph{Analysis for specific networks}

Regarding the results of Section~\ref{sec:adoption-analysis}, it would
be interesting to see if the negative results can be alleviated by
studying special cases of networks. One example is to find classes of
graphs for which we can have efficient constant factor approximation
algorithms for the MIN-ADOPTION problem. We are also not yet aware if
the same hardness results hold for equitable networks. Finally, it
would be interesting to study such problems for graphs that resemble
real networks with respect to degree distribution or other graph
theoretic properties.
 
\paragraph{Optimizing the spread of a product}
Given a diffusion model, one important problem, especially in the
context of viral marketing is: given a network $(G, P,p,\theta)$, a
product $t\in P$, and $k\geq 0$, we wish to find the optimal set
$S$ of nodes, under the restrictions that $|S|\leq k$ and $t\in p(i)$ for $i\in S$, such that
if we give the product $t$ to the members of $S$, optimal spread is
achieved.

The parameter $k$ indicates a bound on the budget for the company's advertising campaign.  
This problem was initially studied
for the case of a single product, and when the thresholds are random variables in~\cite{KKT03} (as noted in~\cite{KKT03}, when thresholds are fixed numbers strong inapproximability results hold).
For the threshold model, some extensions to the case of two products appeared 
in~\cite{BFO10}, where various options on how nodes decide when choosing between two
products have been proposed. The algorithmic challenge here is to compute the optimal set given that the opponents have already chosen their seed sets.
In most cases, the
techniques of~\cite{KKT03} cannot be applied and algorithmic results
are still elusive. It would be interesting to make further progress on this for multiple products.

\section*{Acknowledgements}
The \textbf{MIN-ADOPTION} and \textbf{MAX-ADOPTION} problems from
Section \ref{sec:adoption-analysis} were suggested to us by Berthold
V\"{o}cking. The second author was supported by the Basic Research
Funding Program (BRFP) of the Athens University of Economics and
Business, by the Netherlands Organisation for Scientific Resarch
(NWO), and by the funding program Thales (project ALGONOW, co-financed
by the European Social Fund-ESF and Greek national funds).

\bibliography{e1}
\bibliographystyle{abbrv}

\end{document}